**Multi-frequency coherent emission from superstructure thermal emitters**

*Guanyu Lu, Marko Tadjer, Joshua D. Caldwell\* and Thomas G. Folland\**

Guanyu Lu, Prof. Joshua D. Caldwell, Prof. Thomas G. Folland

School of Mechanical Engineering, Vanderbilt University, Nashville, TN, 37212

E-mail: Josh.caldwell@vanderbilt.edu

Dr. Marko Tadjer

U.S. Naval Research Laboratory, 4555 Overlook Place, Washington DC, 20375

Prof. Thomas G. Folland

School of Physics and Astronomy, The University of Iowa, Iowa City, Iowa, 52242

Email: thomas-folland@uiowa.edu

Keywords: nanophotonics, surface phonon polariton, thermal emitters, coherent emitters, optical design

**Abstract**

Long-range spatial coherence can be induced in thermal emitters by embedding a periodic grating into a material supporting propagating polaritons or dielectric modes. However, the emission angle and frequency cannot be defined simultaneously and uniquely, resulting in emission at unusable angles or frequencies. Here, we explore superstructure gratings (SSGs) to control the spatial and spectral properties of thermal emitters. SSGs have long-range periodicity, but a unit cell that provides tailorable Bragg components to interact with light. These Bragg components allow simultaneous launching of polaritons with different frequencies/wavevectors in a single grating, manifesting as additional spatial and spectral bands upon the emission profile. As the unit cell period approaches the spatial coherence length, the coherence properties of the superstructure will be lost. Whilst the 1D *k*-space representation of the grating provides insights into the emission, the etch depth of the grating can result in strong polariton-polariton interactions. An emergent effect of these interactions is the creation of polaritonic band gaps, and defect states that can have a well-defined frequency and emission angle. In all, our results show experimentally how even in simple 1D gratings there is significant design flexibility for engineering the profile of thermal emitters, bound by finite coherence length.

# 1. Introduction

Coherent thermal emission from periodically patterned surfaces[1] has been exploited for numerous applications from non-dispersive infrared (NDIR) spectroscopy[2, 3], to thermophotovoltaics[4]. Such sources are relatively simple, relying only on a local heat source for light generation, yet can produce significant amounts of emitted power in the mid-infrared. This overcomes some of the limitations of semiconductor optoelectronics for narrowband mid-infrared light emitters, which can be expensive and become increasingly less efficient when operating at longer wavelengths. Coherent thermal emitters rely on a surface-confined wave that propagates along the periodically structured material in the near-field, thereby inducing far-field spatial coherence via coupling to grating modes. One route to realizing such coherent thermal sources has been via exploiting surface polaritons, which are evanescent waves that propagate along the interface between an optical metal and a dielectric (e.g. metal surface in air). Surface polaritons are induced through light coupling with an oscillating charge, with two of the predominant forms within the infrared resulting from coupling to free electrons or polar optic phonons, forming surface plasmon (SPPs) and surface phonon polaritons (SPhPs), respectively[5, 6]. While both are characterized by an evanescent field that propagates along a surface, in the infrared SPPs generally possess frequency tunability and short lifetimes (broad spectral linewidths), whereas SPhPs offer limited frequency tunability and long lifetimes (narrow linewidths)[6, 7]. One of the first demonstrations of spatially coherent thermal emission employed a patterned silicon carbide surface[8, 9], but such efforts have now been extended to include plasmonic structures[10-12] and highly dispersive dielectric resonanators[13]. Yet some of the best performance metrics achievable by periodic structures have been realized through the use of simple 1D gratings[11] or bullseye designs using metals[10], demonstrating the potential for spectrally and spatially narrow emission. However, one of the limitations of 1D periodically patterned gratings is that there is a continuous relationship between incident angle and frequency of the coherent mode. Therefore, all frequencies of light in the operational band are emitted but distributed over different angles based on the grating condition. Further, emission occurs only at normal incidence for bullseye structures, again limiting the application space for periodic thermal emitters.

More sophisticated control of light propagation and advanced functionality can be induced through interference of light waves with different wavevectors. This forms the basis of holography, where an inverse-designed amplitude and/or phase mask is used to create images in the far-field[14]. A number of different structures have been proposed and demonstrated for tailoring the interference of light[15], which can be classified by the type of order or disorder they exhibit. We can describe structures that have short-, but no long-range order (for example chirped gratings[16]), long-, but no short-range order (e.g. SSGs[17], sampled gratings[18], and quasi-crystals[19]), or alternatively no order (as in holograms[20, 21]). All these structures have many more degrees of design freedom than a periodic grating, and yet obtaining a desired output is challenging due to the large combination of parameters that can be chosen. Typically, a design methodology either relies on approximate analytical results, or has a huge computational cost to sample at least some of the design parameters. The development of robust inverse-design techniques in recent years has presented a solution to this problem, with examples drawing from holography[21], machine learning[22] and optimization approaches[23-25]. Such concepts have even been applied to the design of thermal emitters for the purpose of improved control of their spectral and spatial emission properties[24, 26]. These arguments present a compelling picture for the demonstration of thermal emitters with embedded advanced functionalities through local structuring, such as lenses or holograms. However, existing studies on more complicated structures are largely theoretical, and do not explicitly describe the role that the finite spatial coherence of the optical modes involved play in this design. Such considerations are critical for selecting an appropriate design space for a given inverse design problem. Most notably, 1D silicon carbide gratings show coherence lengths commensurate with tens of free-space wavelength cycles[8], which strongly influences the thermal emission performance.

Here, we study the spectral and spatial thermal emission profiles of SSGs fabricated into a semi-insulating 4H-SiC substrate. In an SSG, the structure of the unit cell is altered to form an extended 'super' period (*L*), with examples provided in **Figure 1**. Illustrated are two types of superstructure, with a more complex, but repetitive unit cell (design approaches described in the figure caption). The more complex unit cell structure will lead to interactions with light at a wider range of frequencies, and therefore in principle multi-angle emission (**Figure 1a**). The behavior of the SSGs can be understood by calculating the Bragg vectors ($G_i$) of the grating. We know that for a given periodic grating we can describe the coupling from a free-space wave (with wave vector $k_0=2\pi/\lambda_0$ and incident angle $\theta$) to a propagating polariton mode ($k_p$) as;

$$k_p = G_n \pm k_0 \sin(\theta) = \frac{2\pi n}{\Lambda} \pm k_0 \sin(\theta) \quad (1)$$

where *n* is an integer representing the mode order. To extend this approach, we can study the Fourier decomposition of a structure with a permittivity profile $\varepsilon(x)$ as:

$$\varepsilon(x) = \sum_{n=-\infty}^{\infty} \varepsilon_n e^{inG_x x}. \quad (2)$$

As such, we can establish that an arbitrary structure with a 'super period' *L* will therefore have Bragg components $G_i = nG_x$, with a relative amplitude given by $\varepsilon_n$ where $G_x=2\pi/L$. This 1D Fourier decomposition (or *k*-space profile) forms the basis of the Fourier modal method[27], and provides the general trends of the thermal emission response for shallow gratings[28]. We show how a change in the grating structure has a consequence on the Fourier components in **Figure 1** (calculated from the discrete Fourier transform of the grating profile). All gratings shown possess a Bragg component at the frequency of the same base periodic grating. However, the SSGs also possess additional Bragg vectors, which can couple light into and out of free space for inducing and potentially dictating the spatial coherence. We will show that these Bragg components manifest as additional spatially coherent modes in the normally incoherent thermal emission response, occurring at a different frequency and angle with respect to the primary Bragg wave vector. This results in a multi-frequency, coherent thermal emitter, where the spectral response can be tailored by the additional degrees of freedom the SSG offers in the design of its grating profile (the repetitive unit cell). Our results also indicate that the limitation on this design approach is confined to SSG periodicities that are shorter than the polariton coherence length. This result provides a limit on the maximum system size for inverse designed emitters, which could have more complex properties than those described here. Finally, we show that depending on the strength of the interactions between SPhPs propagating in different directions, it is possible to engineer a SSG where the forward- and reverse-propagating diffracted waves interfere. The consequence is destructive interference of the emitted waves, and the creation of a polaritonic band gap similar to what has been observed between harmonics in periodic structures[11, 29]. By further modifying the structure, a 'gap' emissive state that only couples to free space at a specific frequency and angle, is observed. Whilst generalizing this result would require the use of inverse design approaches, this work demonstrates that arbitrary control of the emission frequency and angle is possible in thermal emitters with a finite coherence length, with 2D SSGs providing additional flexibility in the design of the emission properties within a full hemisphere. As such, our work demonstrates that an appropriately inverse designed SSG might present an effective way of realizing complex optical fields from thermal sources.

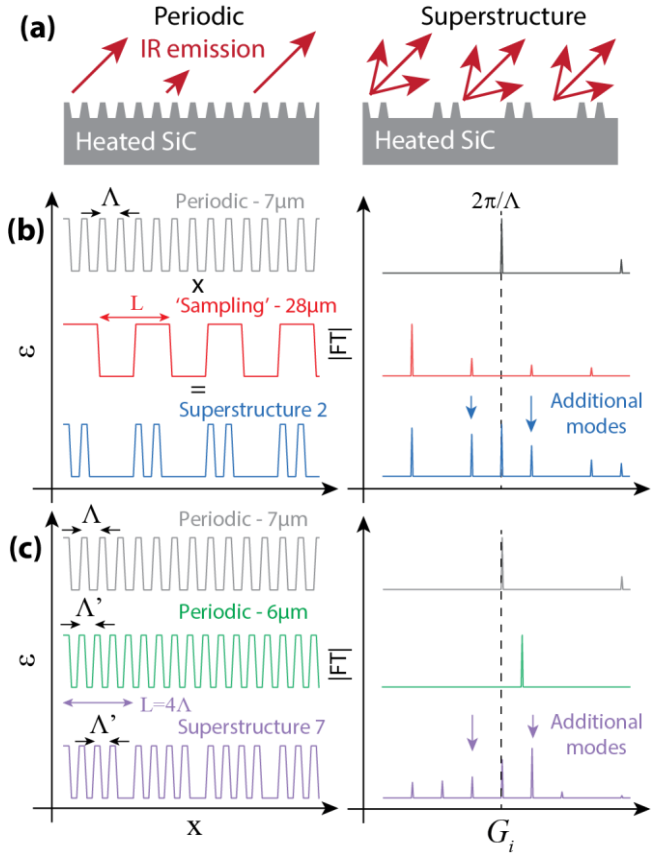

**Figure 1**. Structure and Fourier spectrum of SSGs, for creating multi-angle thermal emission. (a) shows a schematic comparing the emission from a periodic grating to that from a superstructure – illustrating the presence of additional spatial modes. Two types of superstructure grating designs are included in this paper. In (b) we show the creation of a 'sampled grating' superstructure, where a long period square wave is multiplied with a short period wave. The result is a series of 'short' grating elements. In (c) we show a second approach, where the superstructure period L is chosen (with L=NΛ and N is an integer). The unit cell is then constructed of N times a shorter period grating (Λ'), with empty space filling the distance NΛ' to L.

## 2. Methods

To test our hypothesis, a series of 5-mm x 5-mm area SSGs featuring 1-µm-tall grating ridges were fabricated using photolithography and reactive ion etching. Each element of the grating consisted of 2.7-µm-wide grating ridges (**Figure 2a**). All gratings were designed to exhibit a primary $G_i$ at Λ=7 µm using two different forward-design philosophies. The design of the grating structures (**Figure 1b** SSG 1-5, all structures shown in S1) was performed using the traditional 'sampled grating' forward-design methodology widely used in semiconductor laser design[18]. In brief, a periodic grating with a period Λ is multiplied by a square wave with an amplitude of zero to 1 over a period *L*. The result is a series of isolated sections of a periodic grating, with half the total elements of a uniform periodic structure. We also designed structures by taking the period L=NΛ, and then constructing the unit cell from N periods of a short periodic structure (Λ'), with empty space filling the distance NΛ' to L (**Figure 1c**, SSG 6 and 7). This will have the same number of elements as a periodic grating. Note that these forward-design approaches were only employed for ease of design and fabrication, in principle an inverse approach would be much better suited for application-specific designs. The thermal emission from these gratings was measured with the SSG heated to 266°C (**Figure 2a**) using a custom-built, angle-resolved thermal emission setup (more completely described in Reference [2]). In brief, a 1" parabolic mirror with a focal length of 6" (angular FW ~ 4°, calibrated using a HeNe laser) collects light from a sample mounted on a rotatable (emitter angles between 0 and 70°) hotplate. This emitted light is passed through a polarizer (set to collect *p*-polarized light), which is then directed into an FTIR spectrometer (Bruker Vertex 70V, 2 cm$^{-1}$ resolution) equipped with an MCT detector. Three pinholes are included in the optical path to reduce the amount of background IR radiation reaching the detector. In our technique we measure calibrated directional emissivity $\varepsilon_Q$ as a function of angle, relative to a vertically aligned carbon nanotube reference sample that was previously calibrated to have an emissivity of 0.97. Additionally, for comparison we provide the thermal emission spectra of a reference single period grating with 2-µm-tall ridges and Λ=7 µm in **Figure S2**.

## 3. Results and discussion

Angle-dependent thermal emission spectra of SSG1 are represented in the contour plot provided in **Figure 2b**. We observe a spatially coherent mode that disperses as a function of angle in the spectral range of 797 to 870 cm$^{-1}$, in line with numerous earlier experiments[8, 9]. To verify that this mode can be attributed to the existence of a SPhP mode, we can overlay the analytical solution to the grating equation (**Equation 1**) associated with the multiple Bragg vectors present ($G_i$). Instead of presenting these vectors as reciprocal space values, we instead choose to present them in terms of the equivalent grating period $\Lambda$ for clarity. We consider both the forward-propagating grating mode, which will have a positive dispersion with angle, and the reverse-propagating mode that will exhibit a negative dispersion. The spatially coherent mode for SSG1 closely matches the reverse propagating $\Lambda=7$ µm mode from the grating equation with a slight detuning attributed to the finite height of the grating. We note for SSG1 we also anticipate contributions from the $\Lambda=14$ µm ($2^{nd}$ order diffraction) forward-propagating wave, which only manifests above 20° at higher frequencies >900cm$^{-1}$. The forward-emitting wave also interferes with the reverse propagating wave, creating a frequency band with no coherent emission (akin to a photonic band gap). The role of this interference will be discussed later in the paper. For the moment, we instead compare this behavior to that of the response of SSG3 and 5, which both have two secondary Bragg peaks close in wavevector to the primary $\Lambda=7$ µm mode. The emission from these two structures should therefore exhibit multiple dispersive modes, as shown in **Figure 2c** and **Figure 2d**. These new frequency-dispersive modes exhibit a reduced emission intensity in comparison to the primary mode, as the corresponding $G_i$ has a reduced amplitude (**Figure 1b**), leading to a decrease in coupling to free space. Further, each mode matches reasonably well with the associated 'effective periods' extracted from the Fourier spectrum for these components; that is $\Lambda=$ 6 and 8.6 µm for SSG3 and $\Lambda=$ 6.33 and 7.7 µm, respectively. These additional modes are also observed in SSG2 and 4 (shown in **Figure S3**). This demonstrates that SSGs can indeed support more than one coherent, emissive mode enabling the potential for design of the thermal emission spectra and angular distribution.

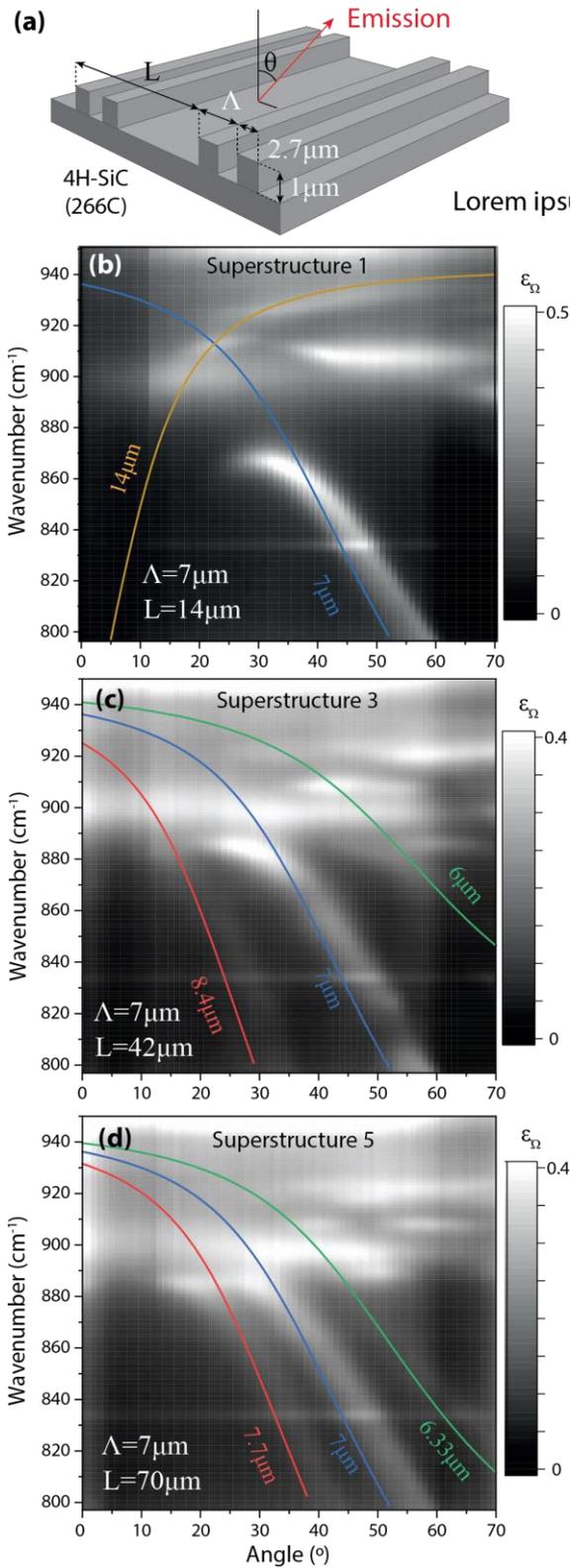

**Figure 2**. Device schematic (a) and frequency and angular dependent emissivity for (b) SSG1, (c) SSG 3 and (d) SSG 5. Overlaid on the plots are the predicted mode frequencies assuming the grating couples to a propagating surface polariton mode. We observe that for superstructure 3 and 5 two side modes appear in the emissivity, corresponding to the additional superstructure modes

The differences in the behavior of the devices can be more clearly compared by taking a single frequency and plotting the angle-resolved emission from the device, as shown in **Figure 3a**. We observe that the side 'lobes' on the emission change significantly with design, but the primary emission frequency shows minimal tuning compared with the measured angular spread of approximately 9°. For further analysis, and comparison to the *k*-space behavior of the grating, we can convert the angular data of **Figure 2d** to wavevector units by using eq. (1). We then normalize the data to the primary $\Lambda=7\mu m$ emission line through an x-axis shift applied to each frequency independently. The purpose of this is to overlay the Fourier spectrum of **Figure 1** more effectively on top of our measured spectra, as shown for SSG5 in **Figure 3b**. We see that the secondary emission peak with positive wavevector (labelled upper band [UB]) lines up extremely well with the secondary Bragg component. The mode featuring the negative wavevector (labelled lower band [LB]) does not line up as closely, however, this is largely limited to the maximum collection angle of our variable-angle system. To further verify that these additional modes are due to the SSG pattern, we can also compare the angle-resolved data collected between SSGs 1 through 5, using the same normalization process, as shown in **Figure 3c**. We see that SSG2 through 5 exhibit a secondary emission peak with positive wavevector (labelled UB) that lines up well with the spacing predicted from the Fourier transform of the grating response. This provides a complete demonstration that superstructure gratings can indeed be used to induce additional emission frequencies into the response of a grating-based thermal emitter.

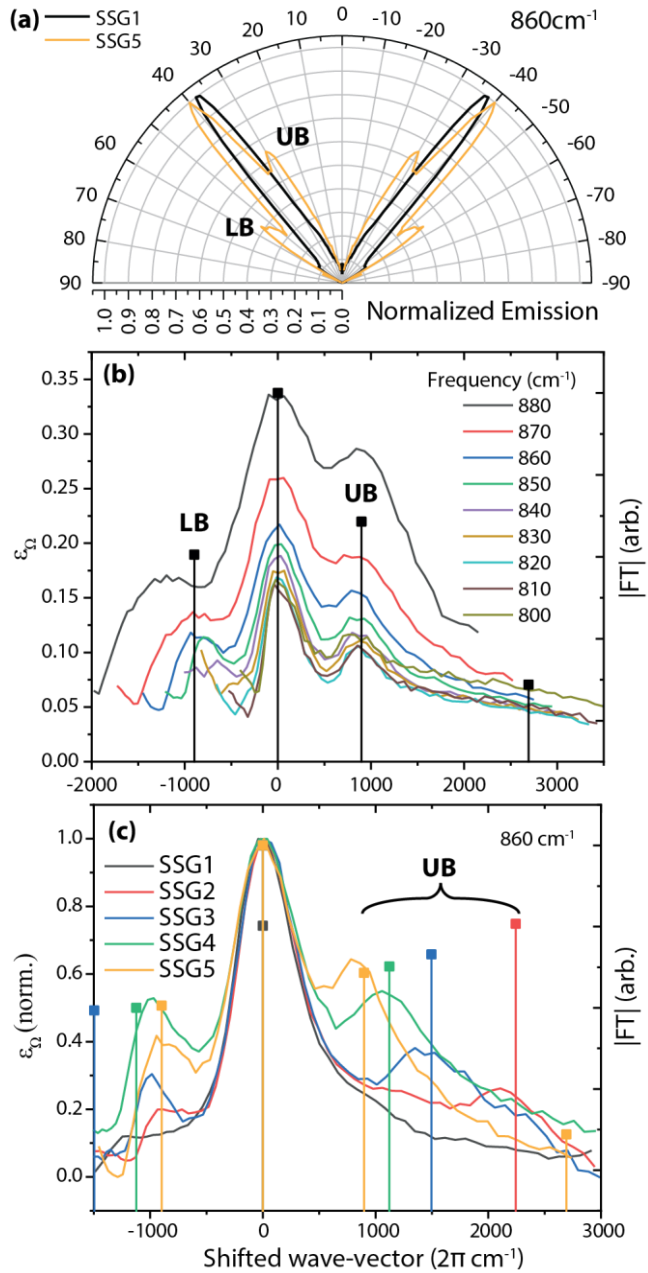

**Figure 3**. Comparison of emissivity vs grating wave vector components for different frequencies and structures (a) shows the emission profile of SSG samples as a function of angle (n) shows the frequency dependent emission of SSG5 normalized to wave vector instead of angle, showing both an upper band (UB) and a lower band (LB) (c) shows the comparison of the emissivity spectrum from different sample at a frequency of 860cm$^{-1}$. The results show that it is possible to engineer multiple spatially coherent modes from a spatially coherent emitter.

One of the obvious questions that arises from this analysis is the maximum possible SSG period that we can use in our design and still get defined emissivity peaks. We already see from SSG5 (**Figure 3a**) that the emissivity of the UB and LB merge into the central peak at higher frequencies. This behavior should be associated with the relationship between the grating period and the mode coherence length $l_c$ can be defined using the following expression[8, 9]:

$$l_c = \frac{\lambda_0}{\Delta\theta} \qquad (3)$$

where $\Delta\theta$ is the angular spread. By fitting a Lorentz curve to the data in Fig. 3 we extract $\Delta\theta = 9.2°$, at 860 cm$^{-1}$. If we account for the angular spread of the collection optic (assuming a FW 4° box-car spread), we estimate the true angular spread for the primary peak as 8.3° through numerical deconvolution. Using this fitted value, we get a coherence length of $l_c$=81 µm. This makes the super period of SSG5 close to this coherence length (as L=70 µm), suggesting that this value provides a useful tool for assessing the largest possible superstructure that can be suitably designed for a given system. At other wavelengths, the coherence length is extended or reduced due to reduced polaritonic losses, and the three modes are more or less defined, respectively.

To assess whether the superstructure approach works on samples with a different design methodologies we also fabricated gratings SSG6 and 7. As described above, these were designed by taking the period L=NΛ, and then constructing the unit cell from N periods of a short periodic structure (Λ'), with empty space filling the distance NΛ' to L. In real space, these gratings have twice the number of grating 'elements', and so appear physically quite different. However, these also exhibit a series of $G_i$ modes closely spaced around the primary 7-µm pitch in *k*-space (**Figure 1c**). Our experimental results for SSG6 show that towards the lower frequency edge of the SiC Restrahlen band (782 cm$^{-1}$), the coherent mode is still supported, tuning with angle (**Figure 4**). This mode matches with the solution to the grating equation for $\Lambda$=7µm, as shown with the blue solid line. However, we also note that for a much wider range of frequencies (approx. 825 to 880 cm$^{-1}$) no coherent mode is observed.

This is not in line with a loss of spatial coherence, as our experiments in **Figure 2** and **Figure 3** have much longer SSG periodicities ($L$=70 µm for SSG5 versus $L$=21 µm for SSG6) and still exhibit such a coherent mode. Instead, we note that the forward-propagating mode attributed to the component at $\Lambda$=21 µm and $\Lambda$=7 µm cross each other in the dispersion relation within the middle of the Reststrahlen band. As these are waves with similar wavevectors, but opposing propagation directions, these can interfere, thereby introducing a photonic band gap. Indeed, earlier work on periodic gratings have noted that in the case of significant scattering from individual grating elements, a polaratonic band-gap (PBG) can open up[11, 29]. This hypothesis is further confirmed by measurements on SSG7, which shows an even wider photonic gap than SSG6, again through the interference of two grating components. This suggests that polariton-polariton scattering plays a significant role in the spectral and spatial properties of both SSG6 and 7.

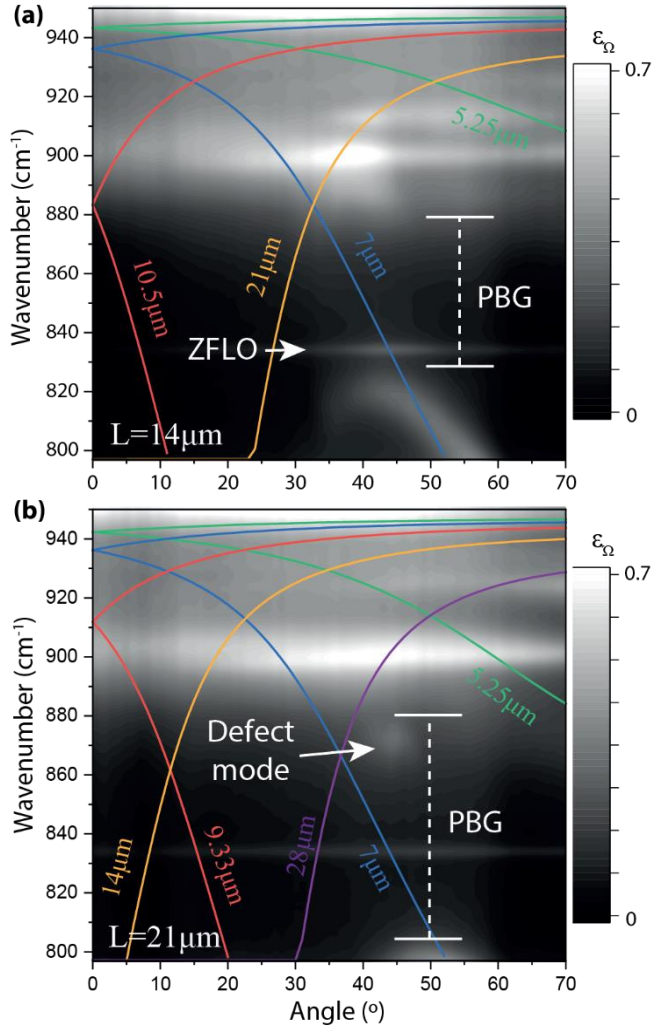

**Figure 4.** Thermal emission using a second superstructure design approach (a) and (b) show the angle resolved emissivity from SSG6 and SSG7 respectively, highlighting the presence of a polaritonic band gap (PBG) and an isolated defect mode within the superstructure spectrum, as well as the zone-folded LO phonon (ZFLO) of 4H-SiC. Overlaid on the plots are the predicted mode frequencies assuming the grating couples to a propagating surface polariton mode.

The origin of these interactions can be understood by simulating a range of different grating depths, as shown in **Figure 5**. Simulations were conducted using CST studio suite 2020 by calculating the absorption of SSG 6 as a function of light frequency and angle (which can be related to emissivity via Kirchhoff's law). We find that the results for shallow gratings are comparable with earlier experiments on 4H-SiC gratings, and the PBG is not present[11]. Deep gratings show the widest PBG, and we can note the excellent agreement between Fig. 4(a) and 5(d), with some of the smaller features not observed at higher frequencies, likely due to fabrication imperfections. The relationship between this PBG and thickness can be attributed to a reduction in polariton-polariton interference effects, as scattering from each element of the grating gets weaker with a reduced grating height. This also suggests that these destructive interference effects will be significantly weaker for SSG1-5, as they have half the number of grating elements, and hence less opportunities to induce such polariton scattering. Obviously, the inclusion of polariton-polariton interactions will make the design of coherent thermal emitters more challenging than in earlier works that considered only grating modes. This is because it is no longer sufficient to consider only the coupling from free space, which is well described by the grating equation, but also must include such interference effects as well. However, such strong interactions also allow the production of emergent phenomena not seen in a conventional periodic gratings. For example, in **Figure 4b** we can observe that SSG7 supports a mode at approximately 870 cm$^{-1}$ and 45 degrees that is both spatially and temporally non-dispersive (e.g., has a well-defined angle and frequency). We label this as a possible 'defect mode' in the results. This is distinct from the non-dispersing zone folded LO (ZFLO) phonon mode, which occurs at approximately 833cm$^{-1}$, and is a material property of SiC. Whilst the existence of this defect mode at this specific frequency cannot be directly predicted by grating theory, this provides evidence that prior theoretically proposed omnidirectional structures will be experimentally realizable[23, 24]. Such devices likely work on the creation of multiple-scattering events between isolated modes, but proof of this concept is beyond the scope of this paper. However, we note that by using inverse design approaches, such structures can correctly account for the superposition of all possible waves and create such modes at user-defined frequencies and angles.

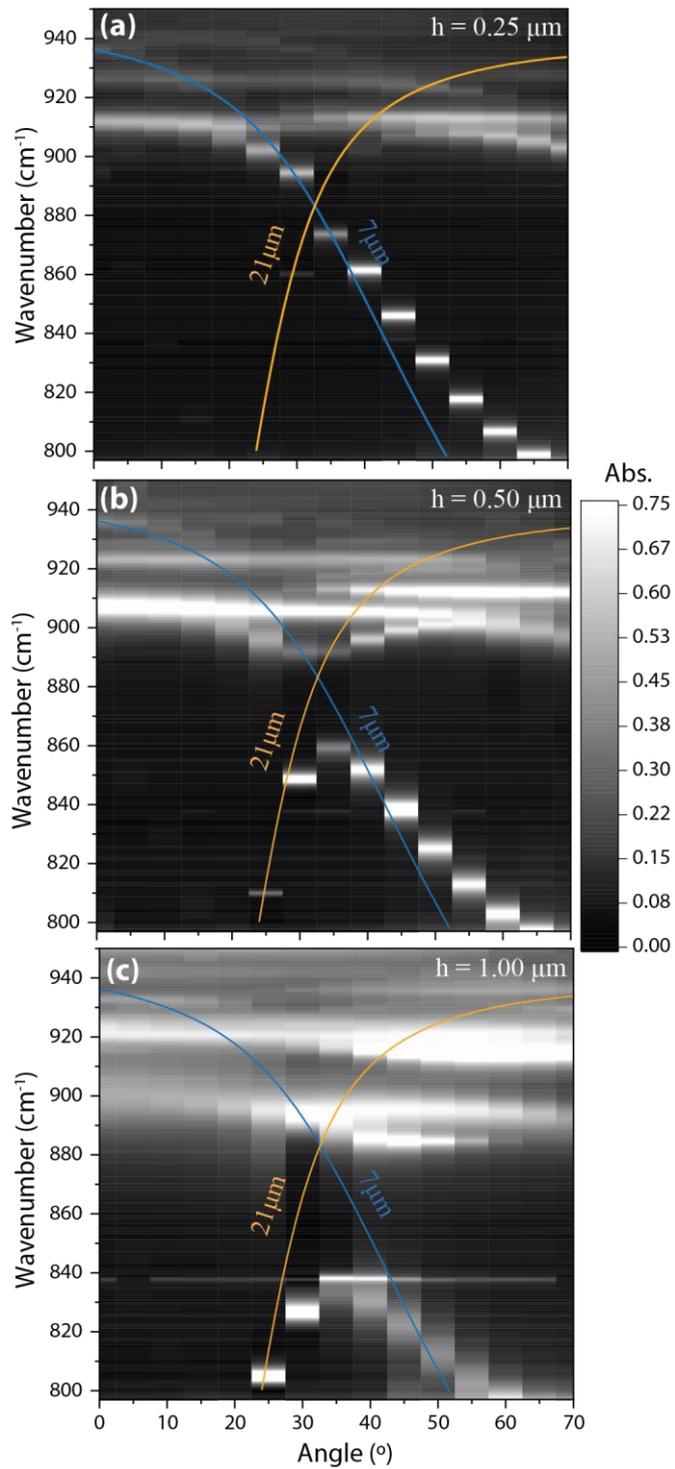

**Figure 5.** Simulated p-polarized absorption for the design of SSG 6 with a etch depth of (a) 0.25 μm, (b) 0.50 μm, (c) 0 cd) 1.00 μm. Calculated forward propagating dispersion (yellow line) and reverse propagating dispersion (red line) are plotted on top of those simulated results. For shallow grating (a), there is almost no interference between the forward and reverse propagating surface waves. Further increasing the etch depth will contribute to the polariton-polariton interactions, thus, the interference between the forward and reverse propagating waves will be more pronounced.

## 4. Conclusion

In this work we have studied how SSGs can be used to tailor the thermal emission spectrum of polaritonic thermal emitters. Our devices exhibit multiple coherent modes existing at different combinations of frequency/angle, controlled by the superstructure design. These correspond to the different diffractive modes revealed by the Fourier transform of the 1D grating profile. The major limitation of this design approach is that the superstructure period must be shorter than the spatial coherence length of the propagating mode to be effective. Further, if the grating has many elements or is etched more deeply into the substrate, polariton-polariton scattering results in the formation of polaritonic band gaps. This induces a deviation from the behavior predicted from the simple grating equation. However, this can also be advantageous, resulting in the formation of a polaritonic band gap, as well as in the potential realization of defect states with a well-defined frequency and angular direction. Whilst the design rules incorporating free space coupling and polariton scattering is beyond the scope of this paper, our work shows the significant potential for spatial engineering of thermal emitters through such SSG design methodologies. Ultimately, we believe that this promises to lead to the creation of sophisticated thermal sources, such as thermal lenses and holograms that offer applications in non-dispersive spectroscopy and/or thermal signature management.

**Supporting Information**
Supporting Information is available from the Wiley Online Library or from the author.

**Acknowledgements**

T.G.F acknowledges funding provided through the School of Engineering at Vanderbilt University through the startup package of J.D.C. J.D.C acknowledges support from the Office of Naval Research under grant number N00014-18-12107. Research at the Naval Research Laboratory was supported by the Office of Naval Research. A portion of this research was conducted at the Vanderbilt Institute of Nanoscale Science and Engineering. Funding for G.L. was provided through a STTR program provided by the National Science Foundation, Division of Industrial Innovation and Partnerships (IIP) (award #2014798)